\documentclass[epjCONF]{svjour}

\usepackage{graphicx}
\usepackage[varg]{txfonts} 
\usepackage[latin1]{inputenc}
\usepackage{longtable}
\def\lsim{\mathrel{\rlap{
\lower4pt\hbox{\hskip-3pt$\sim$}}
\raise1pt\hbox{$<$}}}     
\def\gsim{\mathrel{\rlap{
\lower4pt\hbox{\hskip-3pt$\sim$}}
\raise1pt\hbox{$>$}}}     
\def\be{\begin{eqnarray}}
\def\ee{\end{eqnarray}}
\def\prt{\partial}
\session-title{Hot and Cold Baryonic Matter -- HCBM 2010}
\begin{document}
\title{Energy and system-size dependence of the Chiral Magnetic Effect}
\author{V.Toneev\inst{1}\inst{2}\fnmsep
\thanks{\email{toneev@theor.jinr.ru}} \and V.Voronyuk\inst{1}\inst{3}\inst{4} }
\institute{Joint Institute for Nuclear Research,
141980 Dubna, Moscow Region, Russia \and GSI, Helmholtzzentrum f\"ur
Schwerionenforschung GmbH, 64291 Darmstadt, Germany \and Bogolyubov
Institute for Theoretical Physics, 03680 Kiev, Ukraine \and
 Institute
f\"ur Theoretische Physik, Universit\"at Frankfurt, 60438 Germany}
\abstract{The energy dependence of the local ${\cal P}$ and
${\cal CP}$ violation in Au+Au and Cu+Cu collisions in a large energy
range is estimated within a simple phenomenological model. It is
expected that at LHC the chiral magnetic effect   will be   about
20 times weaker than at RHIC. At lower energy range, covered by
the low-energy scan at RHIC and future NICA/FAIR facilities, the created
magnetic field strength and energy density of deconfined matter are rather
high providing necessary conditions for the chiral magnetic effect.
However, the particular model for the chiral magnetic effect
predicts that this effect should vanish sharply at energy somewhere above
the top SPS one. To elucidate CME background effects the
Hadron-String-Dynamics (HSD) transport model
including electromagnetic fields is put forward. Importance of new planning
experiments  at LHC and for the low-energy RHIC scan program is emphasized.
} 
\maketitle
\section{Introduction}
\label{intro}
Until now there is no direct experimental evidence for topological
effects in QCD while the existence of nontrivial topological
configurations is a fundamental property of the gauge theory.
Transitions between different topological states occur with the
local violation of the ${\cal P}$ and ${\cal CP}$ symmetry.  Due
to chiral anomaly the interplay  of these topological
configurations with (chiral) quarks results in asymmetry between
left- and right-handed quarks. Such chiral asymmetry coupled to a
strong magnetic field, created by colliding nuclei perpendicularly
to the reaction plane, induces a current of electric charge along
the direction of a magnetic field thereby separating particles of
opposite charges with respect to the reaction plane. Thus, as was
argued in Refs.~\cite{Kharzeev:2004ey,KZ07,KMcLW07,FKW08,KW09} the
topological effects in QCD may be observed in heavy ion collisions
directly in the presence of very intense external electromagnetic
fields due to the ``Chiral Magnetic Effect'' (CME) as a
manifestation of spontaneous violation of the ${\cal CP}$
symmetry.  
First
experimental evidence for the CME identified via the observed
charge separation effect with respect to the reaction plane was
presented by the STAR Collaboration at RHIC~\cite{Vo09}. In this
paper we analyze the STAR data in a simple phenomenological way to
estimate a possibility observing the CME in the larger energy
range, from the LHC to FAIR/NICA energies. We also make a step toward
a dynamical estimate of the CME background based on the nonequalibrium
Hadron-String-Dynamics (HSD) microscopical transport approach \cite{HSD}
including electro-magnetic field.


\section{Phenomenological estimates of the CME}\label{sec:2}
\subsection{Model assumptions}
Qualitatively the CME  may be estimated as follows.

We consider the saturation momentum $Q_s$ as a characteristic scale of
the process~\cite{Kharzeev:2004ey}, so the
transverse momentum of particles $p_t \sim Q_s$. Then the total transverse
energy per unit rapidity at mid-rapidity deposited at
the formation of hot matter is expressed through the  overlapping surface
of two colliding nuclei in the transverse plane $S$
\begin{eqnarray}
\label{energy}
\frac{dE_T}{dy} &\sim& \epsilon \cdot V=\epsilon \cdot \Delta z \cdot S
= Q_s\cdot (Q_s^2 S)~.
\end{eqnarray}
Here the energy density and longitudinal size $\Delta z \simeq \Delta \tau
\simeq 1/Q_s$ are taken in order of magnitude as follows
$\epsilon \sim Q_s^4$ and $\Delta z \sim 1/Q_s$. The value in brackets is
proportional  to the hadron multiplicity
$(Q_s^2 S) \sim dN_{\rm hadrons}/dy$, thus we get
\begin{equation}
\label{mult}
\frac{dE_T}{dy} \sim Q_s \cdot \frac{dN_{\rm hadrons}}{dy},
\end{equation}
which simply tells us that the total transverse energy per unit rapidity
is equal to the total number of produced hadrons per unit rapidity times
their average transverse momentum.

The topological charge (winding number) generated in random motion
during the time $\tau_B$ when the magnetic field is present may be
estimated as

\begin{equation}
\label{def1}
n_w \equiv \sqrt{Q^2}=\sqrt{ \Gamma_{S} \cdot V \cdot \tau_B},
\end{equation}
where $\Gamma_S$ is the sphaleron transition rate. Sphalerons describe
a one dimensional random walk in the topological number space. At weak
coupling, $\Gamma_S \sim \alpha_s^5\ T^4$. Little is known about
the rate of topological charge diffusion at strong coupling
$\lambda = g^2 N_c$; however in SUSY Yang-Mills theory the
diffusion rate of topological charge can be evaluated explicitly
and is given by $\Gamma_S \sim \lambda^2\ T^4$.

The initial temperature $T_0$ of the produced matter at time
$\tau \simeq 1/Q_s$ is proportional to the saturation momentum
$Q_s$, $T_0 = c\ Q_s$. Neglecting the expansion time  and the
corresponding time dependence of the temperature we can roughly get
\begin{equation}
\label{res1}
n_w \equiv \sqrt{Q^2} \sim \sqrt{\frac{dN_{\rm hadrons}}{dy}}
\cdot \sqrt {Q_s \ \tau_B}~.
\end{equation}
Since sizable  sphaleron transitions occur only in the deconfined phase,
the time $\tau_B$ in Eq.~(\ref{res1}) is really the smallest
lifetime between the strong magnetic field $\tilde{\tau}_B$ one and the
lifetime of deconfined matter $\tau_{\epsilon}$:
\begin{equation}
\tau_B = {\rm min}\{\tilde{\tau}_B, \tau_{\epsilon} \}.
\end{equation}

The measured charged particle asymmetry  is associated with the averaged
correlator $a$ by the following relation~\cite{Vol05}:
\begin{eqnarray}
\label{cos}
&&\langle \cos (\psi_\alpha+\psi_\beta-2\Psi_{RP}) \rangle =  \\ \nonumber
&=& \langle \cos (\psi_\alpha+\psi_\beta-2\psi_c) \rangle / v_{2,c}=
v_{1,\alpha} v_{1,\beta} - a_\alpha a_\beta~,
\end{eqnarray}
where $\Psi_{RP}$ is the azimuthal angle of the reaction plane defined by
the beam axis and the line joining the centers of colliding nuclei and
averaging in (\ref{cos})  is carried out over the whole event ensemble.
The second equality in (\ref{cos}) corresponds to azimuthal measurements
with respect to particle of type $c$ extracted from three-body correlation
analysis~\cite{Vol05}, $v_1$ and $v_2$
are the directed and elliptic flow parameters, respectively.  According to
 Ref.~\cite{Kharzeev:2004ey} an average correlator
 $a=\sqrt{a_\alpha a_\beta}$
is related to the topological charge, $n_w$, as
\begin{equation}\label{rel}
a \sim \frac{n_w}{dN_{\rm hadrons}/dy}~.
\end{equation}
This estimate neglects absorption and rescattering in dense matter
responsible for the difference of magnitudes between the same and
opposite charge correlations and so is very rough. Nevertheless,
for the purposes of our estimate, the use of Eqs. (\ref{res1}) and
(\ref{rel}) yields the relation
\begin{equation}\label{rel1}
a \sim \frac{\sqrt {Q_s \tau_B}}{\sqrt{dN_{\rm hadrons}/dy}}~,
\end{equation}
on which our consideration is based.

%

\subsection{Energy dependence of the CME}
\begin{figure}[h]
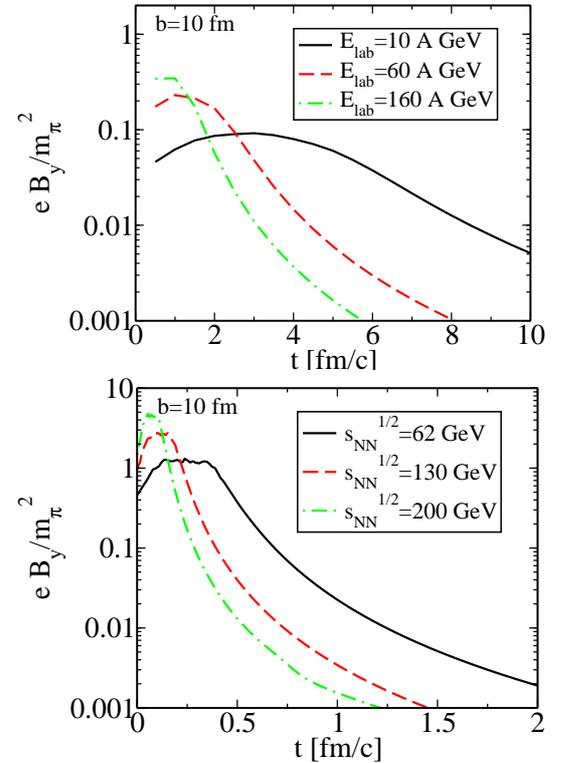

  \begin{center}\resizebox{0.85\columnwidth}{!}{%
 \includegraphics{B_semi_log.eps} }
\resizebox{0.85\columnwidth}{!}{%
\includegraphics{B_semi_log_RHIC.eps} }
\caption{The time evolution of the magnetic field strength $eB_y$
at the central region in  Au+Au collisions  with the impact
parameter $b=10$ fm  for different bombarding energies.
Calculations are carried out within the UrQMD
 model~\cite{Bass:1998ca,Bleicher:1999xi} (for a detail see~\cite{SIT09}).
   \label{B_ev}}
 \end{center}
\end{figure}

\begin{figure}[thb]
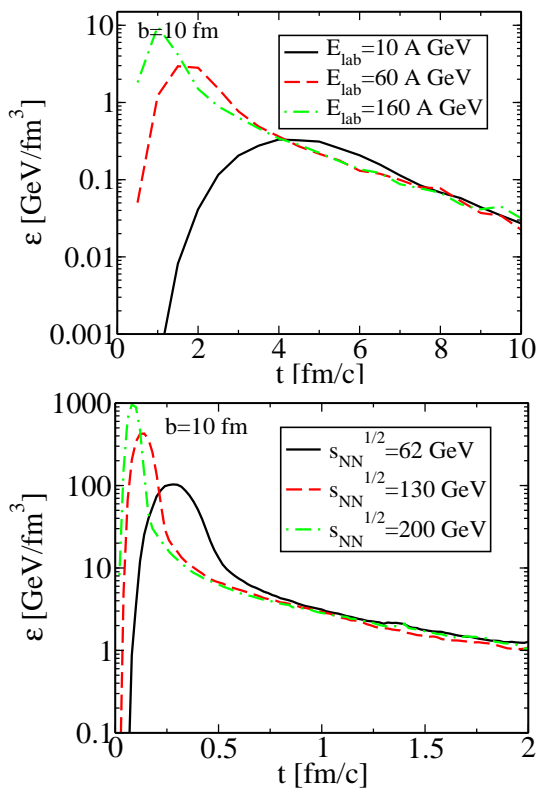

\begin{center}\resizebox{0.85\columnwidth}{!}{%
\includegraphics{E_semi_log.eps} }
\resizebox{0.85\columnwidth}{!}{%
\includegraphics{E_semi_log_RHIC.eps}  }
\caption{ The time evolution of the energy density $\varepsilon$ of created
particles in the Lorentz-contracted box with the 2 fm side
at the central point of overlapping region. The impact parameter $b=10$ fm.
 \label{E_ev}}
 \end{center}
\end{figure}

Let us now come to numerical estimates.
Using the energy dependence $Q_s^2 \sim s_{NN}^{1/8}\sim dN_{\rm
hadrons}/dy$ deduced from RHIC and HERA~\cite{Kharzeev:2001gp,KN01}
data we finally get  for the energy dependence
\begin{equation}
\label{res2}
a \sim \sqrt{\frac{\tau_B}{Q_s}} \sim (\sqrt{s_{NN}})^{-1/16}
\cdot \sqrt{\tau_B}.
\end{equation}
Thus, the direct energy dependence is comparatively weak. Results of
dynamical heavy-ion calculations of the magnetic field at the central
point of the transverse overlapping region of colliding nuclei and energy
density of created particles are
presented in Figs.~\ref{B_ev} and \ref{E_ev}, respectively. Here for a
field estimate we follow Ref.~\cite{SIT09} basing on
the UrQMD model~\cite{Bass:1998ca,Bleicher:1999xi} and applying
the electromagnetic Lienard-Wiechert potential with the retardation
condition for the magnetic field.
As is seen,  at the impact parameter $b=10$ fm
the maximal strength of the dominant magnetic field component $B_y$
(being perpendicular to the reaction plane) is decreased in Au+Au collisions
by the factor of about 10, when one proceeds
from $\sqrt{s_{NN}}=$200 GeV to $E_{lab}=$11 GeV, while for  the  created
particle energy density $\varepsilon$ in the central box this factor is 250
i.e. noticeably higher.

To use Eq. (\ref{rel1}) we need to identify the impact
parameter, saturation momentum and multiplicity at a specific centrality.
These can be found in Ref.~\cite{KN01} where the Glauber calculations were
done. As a reference point we choose $b=$10 fm  in our subsequent
consideration.

The measured value of $\langle \cos
(\psi_\alpha+\psi_\beta-2\Psi_{RP})\rangle$ for the same charge particles
from   Au+Au ($\sqrt {s_{NN}}=200$ GeV) collisions at the impact parameter
$b=$10 fm (40-50$\%$ centrality interval)  is
$-(0.312\pm 0.027)\cdot 10^{-3}$~\cite{Vo09}.
Appropriate number for $\sqrt {s_{NN}}=62$ GeV seems to be a
little bit larger but for Cu+Cu collisions the effect is
definitely stronger~\cite{Vo09}.
Thus, ignoring any final state interactions with medium, assuming $a_\alpha
=a_\beta=a$ and neglecting the directed flow
$v_{1a}=v_{1b}=0$ we get from Eq.~(\ref{cos}) $a^2_{exp}=0.31\cdot
10^{-3}$ for the maximal RHIC energy. Using numbers for the  $\sqrt{s_{NN}}=$200
GeV reference case, from Eq.~(\ref{res2}) we may quantify the $\cal{CP}$
violation effect  by the correlator
\begin{equation}
\label{res3}
a^2  =  K_{Au} \  (\sqrt{s_{NN}})^{-1/8} \cdot \tau_B~.
\end{equation}
The normalization constant $K_{Au}$ can be tunned  at the reference energy
$\sqrt{s_{NN}}=$200 GeV from the inverse relation and experimental
value $a_{exp}$ at this energy for $b=$10 fm
\begin{equation}
\label{res4}
K_{Au}=\frac{ a^2_{exp} \cdot (200)^{1/8} } { \tau_B(200)}~.
\end{equation}
The lifetime $\tau_B$ may be defined as the time during which the
magnetic field is above the critical value needed to support a
fermion Landau level on the domain wall $eB_{crit} = 2 \pi/ S_d$,
where $S_d$ is the domain wall area. Since the size of the domain
wall is not reliably known, it is hard to pin down the number, but
it should be of the order of $m_\pi^2$. Honestly, we have to treat
it as a free parameter.

Indeed the size of the topological defect (say, a sphaleron) in the
region between $T_c$ and 2$T_c$
is very uncertain. At weak coupling, the size is determined by the
magnetic screening mass and it  is $\sim 1/(\alpha_s T)$.
If one plugs $\alpha_s\approx$0.5 and $T =$ 200 MeV, the size is  of
about 2 fm and then the threshold field
is very small $eB_y \sim (\alpha_s T)^2 \sim 0.2 \ m_\pi^2$.

On the other hand, we know that between $T_c$ and $2T_c$ the magnetic
screening mass which determines the size
of the sphaleron is not small as expected from the perturbative theory,
$\alpha_s T$, but from the lattice it is  numerically large till about
5$T_c$. This would increase the
threshold to 20 $m_\pi^2$, however the relation between magnetic mass and
the sphaleron size is valid only as long as the coupling is weak.

All we can say it is perhaps in between (0.2$-$20) $m_\pi^2$. Eventually
lattice QCD calculations may clear this up.

\begin{table}
\caption{
Estimated parameters for the ${\cal CP}$ violation effect in Au+Au
collisions at centrality (40-50)$\%$ with the critical field
$eB_{crit}=0.2 \ m_\pi^2$.
}
\label{tabl2}
\begin{center}\begin{tabular}{|c|c|c|c|c|}
\hline\noalign{\smallskip}
$\sqrt{s_{NN}} \ $GeV & \ $s_{NN}^{1/16}$ \ & $\tilde{\tau}_B$, fm/c& $\tau_\epsilon$, fm/c & $a^2$ \\
\hline\noalign{\smallskip}
$4.5 \cdot 10^3$ & 2.86  & 0.018  & $>$1 &0.016$\cdot 10^{-4}$  \\
200    & 1.94 & 0.24 & $>$2 &0.31$\cdot 10^{-3}$  \\
130    & 1.84  & 0.33& $\sim$2.3 &0.45$\cdot 10^{-3}$ \\
62     & 1.68  & 0.62 & $\sim$2.2  &0.93$\cdot 10^{-3}$ \\
17.9   & 1.43 & 1.41 & $\sim$2. & 2.48$ \cdot 10^{-3}$\\
11. & 1.35 & 1.66&$\sim$ 1.9 &3.10$\cdot 10^{-3}$\\
4.7  & 1.21 & 0. & 0. &0.\\
\hline\noalign{\smallskip}
\end{tabular} \end{center}
\end{table}


The upper bound on the magnetic strength $eB_{crit}=20\ m_\pi^2$
results in $\tau_B=0$  even for the RHIC energy and therefore in this
case the CME should not be observable at all in this energy
range. The time evolution of the magnetic field and energy density,
$\varepsilon$, of newly created hadrons are presented in
Figs.~\ref{B_ev} and \ref{E_ev}. The extracted values of $\tau_B$
defined by the constraints $eB_y>0.2\ m_\pi^2$  and
$\tau_\epsilon$ ($\epsilon >1 \ {\rm GeV} /{\rm fm}^3$) are summed
in Tabl.~\ref{tabl2}. For the reference energy and the  minimal
magnetic field constraint we have
\begin{equation}
\label{res5}
K_{Au}=2.52\cdot 10^{-3}.
\end{equation}
If lifetimes are known for all energies one can estimate the
$\cal{CP}$ violation effect through the $a^2$ excitation function.

>From the first glimpse as follows from Tabl.~\ref{tabl2}, in the case of
$eB_{crit}=0.2 \ m_\pi^2$ the interaction time $\tau_B$
is defined solely by evolution of the magnetic field since
$\tilde{\tau}_B<\tau_\varepsilon$ whereas $\tau_\varepsilon \approx 2$ fm
independent of $\sqrt {s_{NN}}$. The expected CME for Au+Au at $b=10$ fm
(see the last column in Tabl.~\ref{tabl2}) monotonously increases when
$\sqrt{s_{NN}}$ goes down but then sharply vanishes exhibiting
a shallow maximum in the range between near the top SPS and NICA
energies. The position of CME maximum and its magnitude depend on the cut
level which just defines $\tilde{\tau}_B$.
The decrease of the $eB_y$ bound to 0.02\ $m_\pi^2$ shifts the
maximum toward lower  energy $\sqrt{s_{NN}}$ and enhances
its magnitude. In an opposite limit
when  results are extrapolated to the LHC energy, the CME falls down by a
factor of about 20 with respect to the RHIC energy. This result is quite
understandable. The CME is mainly defined by the relaxation time of the
magnetic field which is concentrated in the Lorentz contracted nuclear
region $\sim 2R/\gamma$. Therefore, the CME inversely proportional to the
colliding energy, $\sim 1/\sqrt{s_{NN}}$ and proceeding from the RHIC to
LHC energy we roughly get the suppression factor about 4.5/0.2$\eqsim$ 22.

There is one worrying point here. Proceeding from \\ $\sqrt{s_{NN}}=$200
to 62 GeV the predicted value of $a^2$ for $b$=10 fm increases in three
times though not more 20$\%$ growth has been observed in these collisions
in the recent experiment~\cite{Vo09}. This essential disagreement cannot
be removed by a simple variation of $eB_{crit}$. One may try to explain
this correlator overestimation at $\sqrt{s_{NN}}=$62 GeV by an irrelevant
choice of the energy dependence of multiplicity in Eq.~(\ref{res3}). For
the correlator ratio  at these two energies we have
\begin{eqnarray}
\frac{a^2(200)}{a^2(62)}&=&\frac{\tau_B(200)}{\tau_B(62)}
\left( \frac{62}{200}
\right)^{1/8} \nonumber \\ &=&0.387 \ (0.31)^\beta \approx 0.72.
\label{ratio}
\end{eqnarray}
where we use lifetime values from Tabl.~\ref{tabl2} and experimental values
for correlators~\cite{Vo09}, $\beta\equiv 1/8$. As follows from
Eq.~(\ref{ratio}), to explain the experiment the exponent should be
negative, $\beta <0$. Therefore, the fast growth of $\tau_B$ with the
energy decrease cannot be compensated by uncertainty in energy dependence
of the correlator $a$.

\begin{figure}[h]
\resizebox{0.75\columnwidth}{!}{%
\includegraphics{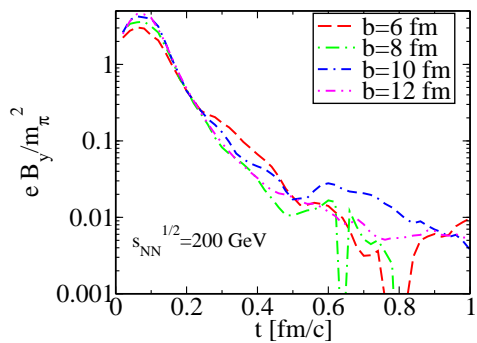} }
\caption{The time evolution of
the magnetic field strength $eB_y$ at the central region
in  Au-Au (200 GeV) collisions for different impact parameters.
\label{b_tau}}
\end{figure}

Uncertainty in the choice of the impact parameter does not help us to solve
this issue as well. As is seen in Fig.~\ref{b_tau} the maximal strength of
the magnetic field slightly increases with the impact parameter but the
lifetime of strong magnetic field $\tilde{\tau}_B$
practically is independent of the impact parameter (for example, look at the
level of $eB_{crit}=0.2 \ m_\pi^2$). It will be used below as an estimate
of the $b$-dependence of the CME which is roughly proportional to
$b$~\cite{KMcLW07}.

As follows from Eq.~(\ref{ratio}), to eliminate model discrepancy
between two RHIC energies the lifetimes should be comparatively
close to each other, $\tau_B(62) \approx 1.2\tau_B(200)$. One may
try to fit this ratio by  the variation of $eB_{crit}$. As is seen
from Tabl.~\ref{tabl2} at $eB_{crit} = 0.2 \ m_\pi^2$ we have
$\tau_B(62)/\tau_B(200)\approx 3.0$. The needed time ratio, 1.2, is
achieved at $eB_{crit}\sim 1.05 \ m_\pi^2$, very close to the
maximal magnetic field of $\sim 1.2 \ m_\pi^2$ for
$\sqrt{s_{NN}}$=62 GeV (see Fig.\ref{B_ev}). It is evident
that if the low-bound field
$eB_{crit}$ coincides with the maximal one the CME should not be
observable. For more accurate tuning new experimental data at
$\sqrt{s_{NN}}\lsim$ 62 GeV are very needed. The decrease of the
critical field till $eB_{crit}\sim 0.01\ m_\pi^2$ makes
$\tilde{\tau}_B$ larger; however, the condition
$\tilde{\tau}_B<\tau_\varepsilon$ which could improved situation
for $\tau_B$ is not yet achieved at $\sqrt{s_{NN}}=$62 GeV.
Unfortunately,  the influence of this lower bound of the $eB_y$
insufficient for manifestation of this effect.

The CME assumes that there are soft equilibrated quarks-gluons.
But it is hard to think that there is an quark-gluon equilibration
in the very early  state and the concept of the initial time is
introduced~\cite{KMcLW07}. It is used to suppose that  an
equilibrium state is achieved at the expansion stage which starts
just after passing the maximum in the energy density evolution.
Thus, the $\tau_\varepsilon$ value should be count off from this time moment
$t_i(\sqrt{s_{NN}})$ rather than from zero. This correction makes
shorter the lifetime by $t_{i,\varepsilon}=$0.08 and $\sim$0.32 fm
for $\sqrt{s_{NN}}=$200 and 62 GeV, respectively, as it follows
from Fig.~\ref{E_ev}. Thus, the lifetime ratio at two energies
discussed is getting less, $\tau_B(62)/\tau_B(200)\approx
(0.62-0.32)/(0.24-0.08)\approx 2$ instead of the mentioned
$\approx$3.0. This softens the dependence of $a^2(200)/a^2(62)$ on
$\sqrt{s_{NN}}$ but it still is not sufficient.

According to Ref.~\cite{KMcLW07} the initial time is estimated as
$t_i\sim 1/Q_s$, where $Q_s\sim s^{1/16}_{NN}$, as noted above.
Normalizing the initial time at $\sqrt{s_{NN}}=$200 GeV to
$t_{i,\varepsilon}=$0.08 fm we get $\tau_B=$0.16 fm for 62 GeV.
This shift is too short to influence essentially on the
$\tau_B(200)/\tau_B(62)$ ratio. With reference to the Lorentz
contraction, other scenario maybe proposed as $t_i\sim
1/\sqrt{s_{NN}}$~\cite{KMcLW07,Ok09}. Under similar normalization
we arrive at $t_{i}\approx$0.27 fm which is rather close to
$t_{i,\varepsilon}$ at 62 GeV in the above estimate and therefore
also is unsatisfactory.

\begin{figure}[h]
\resizebox{0.95\columnwidth}{!}{%
\includegraphics[angle=-90,width=8.truecm] {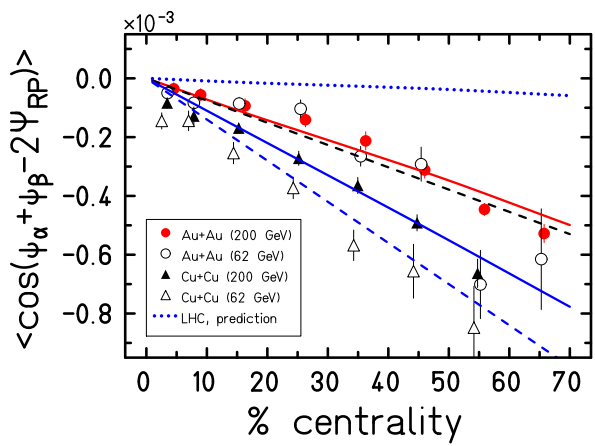} }
\caption{Centrality dependence of the CME. Experimental points for Au$+$Au
and Cu$+$Cu collisions are from~\cite{Vo09}. The dotted line is our
prediction for Au+Au collisions at the LHC energy.
 \label{CME_cu}}
\end{figure}

Now let us try to combine two last scenarios by changing the low bound
of the magnetic field $eB_{crit}$ and simultaneously shifting the
appropriate interaction time to the initial moment defined by
the energy density of created quarks,
$\tau_B=\tilde{\tau}_B(eB_{crit})-t_{i,\varepsilon}$. It is turned
out that the equality (\ref{ratio}) is satisfied if
$eB_{crit}\approx 0.7\ m_\pi^2 $. Using the value of $\tau_B(200)$
obtained in this analysis one can recalculate the coefficient in
Eq.~(\ref{res3}), $K=6.05\cdot 10^{-3}$, and therefore find the
correlator $a$ at any energy. In principle, similar analysis may
be repeated for other impact parameters to consider the
$b$-dependence of the CME. As was shown in
Refs.~\cite{KMcLW07,Ok09} the CME roughly is linear in $b/R$.
Taking this as a hypothesis we evaluate the centrality dependence
of the CME fitting this line to points $b=10$ fm (or centrality
$(40-50)\%$) to be estimated in our model and $b=0$ where the CME
is zero. The results are presented in Fig.~\ref{CME_cu} for Au+Au
collisions at three energies.

As it is seen the calculated lines quite reasonably reproduce the
measured points of azimuthal asymmetry of charge separation for
Au$+$Au collisions at $\sqrt{s_{NN}}=$200 and 62 GeV. The chosen value
of $eB_{crit}=0.7\ m_\pi^2$ results in absence of the CME
at the top SPS energy because  the critical magnetic field practically
coincides with the maximal field at this bombarding energy (see
Fig.~\ref{B_ev}). The CME at the LHC energy is expected to be
less by a factor of about 20  as compared to that at the RHIC energy.
Note that at the LHC energy we applied a simplified
semi-analytical model \cite{SIT09} for magnetic field creation and
assumed $t_{i,\varepsilon}=0$. Thus, we consider this LHC estimate
as an upper limit for the CME.

%
%

\subsection{System-size dependence of the CME}
\begin{figure}[thb]
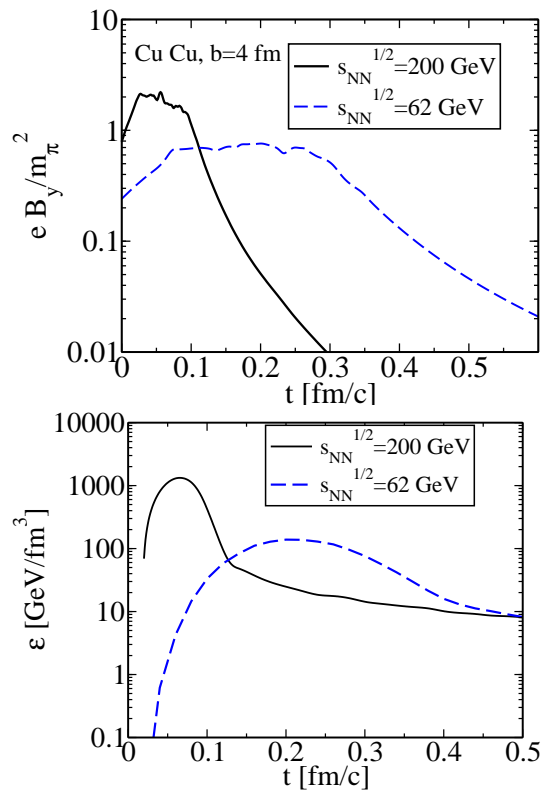

\begin{center}\resizebox{0.85\columnwidth}{!}{%
\includegraphics{B_semi_log_RHIC_cu_62.eps} }
\resizebox{0.85\columnwidth}{!}{%
\includegraphics{E_semi_log_RHIC_cu_62.eps} }
\caption{The time evolution of the magnetic field strength $eB_y$
and deconfined energy density $\varepsilon$ at the central region in  Cu+Cu
collisions  at $\sqrt{s_{NN}}=$200 and 62 GeV for centrality (40-50)$\% \
(b=4.2 \ fm)$.
\label{B_ev_Cu_62} }
\end{center}
\end{figure}

Similar analysis may be repeated for Cu+Cu collisions basing
on available RHIC measurements at two collision energies.
Here one remark is in order.
An enhancement of the CME in Cu+Cu collisions with respect to Au+Au ones
was seen experimentally at the same centrality~\cite{Vo09} but not at the
same impact parameter. As follows from the Glauber
calculations~\cite{Vogt07}, the impact parameter b=10 fm for gold reactions
corresponds to centrality (40-50)$\%$  while the same centrality for copper
collisions  matches  b=4.2 fm. As is seen in Fig.\ref{B_ev_Cu_62} the
time distributions of the magnetic field and energy density for Cu+Cu
collisions look very similar to that for Au+Au ones but  lifetimes, both
$\tilde{\tau}_B$ and $\tau_\varepsilon$, are shorter  in the Cu+Cu case.
The extracted lifetimes and other characteristics for $eB_{crit}=0.2
m_\pi^2$ ($K_{Cu}=6.34\cdot 10^{-3}$) are presented in Tabl.~\ref{tabl3}.
We meet again the same problem: one should compensate a too
strong energy dependence of the model correlators by the proper definition
of lifetimes. Defining the lifetime  in the same manner as for Au+Au
collisions $\tau_B=\tilde{\tau}_B(eB_{crit})-t_{i,\varepsilon}$ the lifetime
ratio  at $eB_{crit}=0.3\ m_\pi^2$ for Cu+Cu collisions
$\tau_B(62)/\tau_B(200)=(0.35-0.21)/(0.14-0.06)$
is turned out to be very close to experimental one. In this case
$K_{Cu}=11.9\cdot 10^{-3}$. In the linear approximation
with the reference point at $b=$4.2 fm, one may draw
the centrality dependence of the CME for Cu+Cu collisions shown also
in Fig.\ref{CME_cu} which is in a reasonable agreement with the experiment.
Note that $eB_{crit}=$0.3 $ m_\pi^2$ which is slightly above the maximal
magnetic field at $\sqrt{s_{NN}}=$62 GeV implies that the CME for Cu$+$Cu
collisions will not be observable even at the top SPS energy.

\begin{table}
\caption{
Estimated parameters for the ${\cal CP}$ violation effect for Cu+Cu
(b=4.2 fm) collisions with $eB_{crit}=0.2 \ m_\pi^2$.
}
\label{tabl3}\begin{center}\begin{tabular}{|c|c|c|c|c|}
\hline \noalign{\smallskip}
$\sqrt{s_{NN}} \ $GeV & $\ s_{NN}^{1/16} \ $  & $\tau_B$, fm/c& $\tau_\epsilon$, fm/c & $a^2$ \\
\hline \noalign{\smallskip}
200  & 1.94  & 0.15 & $>$2.  & 0.49$\cdot 10^{-3}$  \\
62   & 1.68  & 0.37 & $>$2.  & 1.40$\cdot 10^{-3}$ \\
\hline\noalign{\smallskip}
\end{tabular} \end{center}
\end{table}


From dimensionality arguments the system-size dependence of  the chiral
magnetic effect  (at the same all other conditions) would be expected to be
defined by the surface $S\equiv S_{\rm A}(b)$ of an ``almond''-like
transverse area of overlapping nuclei
since  both the high magnetic field and deconfined
matter are needed for this effect. The magnetic field was evaluated
in the the center of the overlapping  region but as was shown
in Ref.~\cite{SIT09} the studied  $eB_y$ component is quite homogeneous
along $y$ of this ``almond''.  Above, say in  the $\sqrt{s_{NN}}$-analysis
of Au+Au collisions, the area $S_{\rm Au}$(b=10 fm) does not depend on
the energy and it is involved into the coefficient $K$, Eq. (\ref{res4}).
Using  for ``almond'' area a  rough estimate as for sharp boundary
overlapping discs of radius $R=r_0 A^{1/3}$, namely
$S\equiv S_A(b)=\pi \sqrt{R^2 - (b/2)^2} (R-b/2)$,  we have
$S_{\rm Cu}(b=4.2)/S_{\rm Au}(b=10)\approx$ 1.65 which seems to be
consistent with experimental ratio of
the CME at these two points. Thus, if one neglects the lifetime on the
system-size and keep only the renormalized coefficient
 $K_{\rm Cu}=K_{\rm Au}\ S_{\rm Cu}(b=4.2)/S_{\rm Au}(b=10)=1.65 K_{\rm Au}$
for the Au curve at $\sqrt{s_{NN}}=$200 GeV in Fig.\ref{CME_cu} we are able
to reproduce rather well the results for Cu+Cu (200 GeV). Note that this Au
curve was obtained  for $eB_{crit}=0.7 \ m_\pi^2$ and non-zero initial
time $t_{i,\varepsilon}$. Thus, conditions
of the compared events are not identical and this success cannot be repeated
for Cu+Cu (62 GeV) collisions where maximum of the magnetic field is lower
than $0.7 \ m_\pi^2$. The above rough estimate essentially depends on
the $r_0$ parameter. From our analysis of the centrality dependence of the
CME at $\sqrt{s_{NN}}=$200 GeV at $eB_{crit}=0.2 \ m_\pi^2$ we have
$K_{\rm Cu}(b=4.2)/K_{\rm Au}(b=10)=6.34\cdot 10^{-3}/2.52\cdot 10^{-3}
\approx 2.5 $ which strongly differs from the above estimate.  Therefore,
the Cu enhancement effect is not only a geometric one. In this case if the
reference point in centrality is considered, the energy dependence should be
taking into account. One should note that in a general case the saturation
momentum $Q_s^2$ is weakly $A$-dependent  as well  what was neglected in
our consideration.

\section{Towards a kinetic approach to the CME background}
\label{sec:3}
The discussed CME signal, the electric charge
asymmetry with respect to the reaction plane,  may originate not
only from the spontaneous local {\cal CP} violation but also be
simulated by other possible effects. In this respect it is important
to consider the CME background. We shall do that considering a full
evolution of  nucleus-nucleus collisions in terms of the HSD transport
model \cite{HSD} but including formation of electromagnetic field as
well as its evolution and impact on particle propagation.

Generalized on-shell transport equations for strongly interacting particles
in the presence of electromagnetic fields can be written as
.

 \be \label{kinEq}
 \{ \frac{\prt}{\prt t}&+&\left(\nabla_{\vec{p}} \ \vec{U}\right)
 \nabla_{\vec{r}}-\left(\nabla_{\vec{r}} \ \vec{U}+{ q\vec{v}\times
 (\nabla \times \vec{A})} \right)\nabla_{\vec{p}} \
 \} \ f(\vec{r},\vec{p},t)\nonumber \\&=&I_{coll}(f,f_1,...f_N)
\ee
 which are supplemented by  equations for magnetic and electric fields
 \be \label{EMf}
\vec{B}=\nabla \times \vec{A}, \ \ \ \ \vec{E}=-\nabla \Phi
-\frac{\prt \vec{A}}{\prt t}~.
 \ee
 The hadronic mean-field $U\sim Re (\Sigma^{ret})/2p_0$ here has an
additional term from the 4-vector electromagnetic potential
$A=\{\Phi,\vec{A}\}$.
The general solution of the wave equations (\ref{EMf}) with charge
distribution $\rho(\vec{r},t)=qn$ and current
$\vec{j}(\vec{r},t)=q\vec{v}$ are
 \be
\label{11.30}  \Phi(\vec{r},t) = \frac{1}{4 \pi} \int
\frac{\rho(\vec{r'},t')\  \delta(t-t'- |\vec{r} - \vec{r'}|/c)}{
|\vec{r} - \vec{r'}|}  \ d^3r ' dt'
 \ee
 for the electromagnetic potential $\Phi(\vec r,t)$ and
 \be
\label{11.31}  {\bf A}(\vec{r},t) = \frac{1}{4 \pi} \int
\frac{\vec{j}(\vec{r'},t') \ \delta(t-t'- |\vec{r} - \vec{r'}
|/c)}{ |\vec{r} - \vec{r'}|}  \ d^3r ' dt'
 \ee
\noindent
 for the vector potential.  For moving point-like charges, one gets
 \be
 \label{11.42}  \rho(\vec{r},t) =
q \ \delta(\vec{r} - \vec{r}(t)); \hspace{0.5cm} \vec{j}(\vec{r},t) = q \
\vec{v}(t)\ \delta(\vec{r} - \vec{r}(t)).
\ee

 \noindent
In this case Eq. (\ref{11.42}) leads to the above used retarded
 Li\'enard-Wiechert  potentials generated by every moving charged
 particle and acting at the point
 $\vec{R}=\vec{r}-\vec{r'}$ at the moment $t$~\cite{SIT09}.
 \begin{figure}[thb]
\begin{center}
\resizebox{0.85\columnwidth}{!}{%
\includegraphics{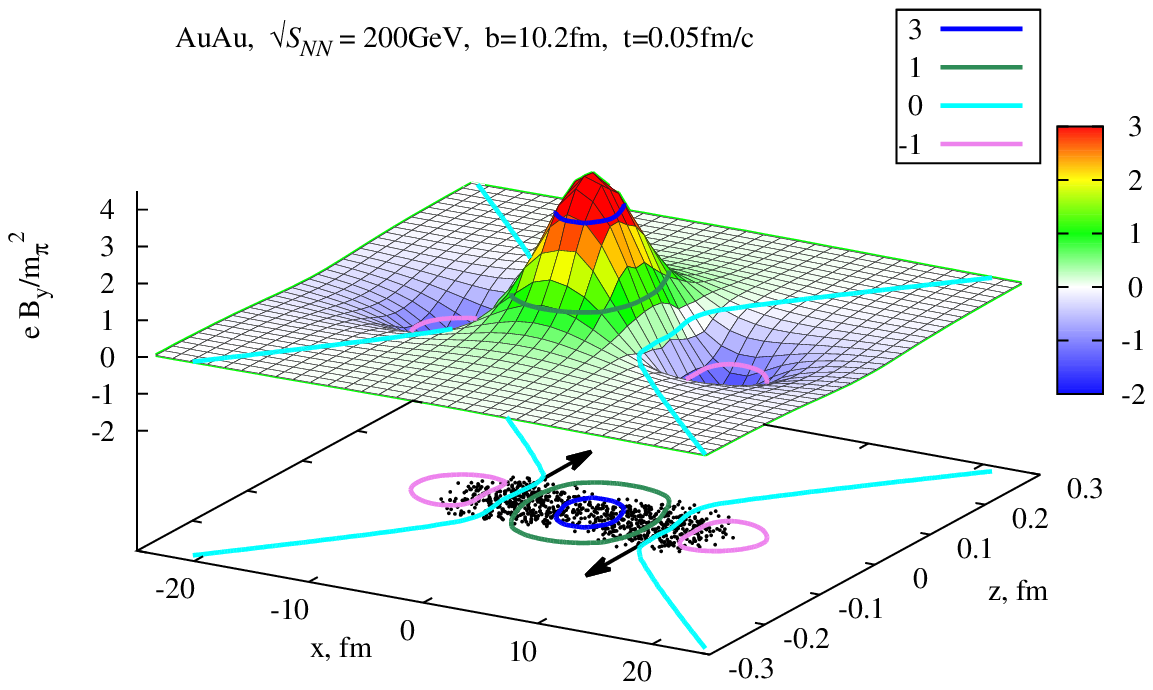} }
\resizebox{0.85\columnwidth}{!}{%
\includegraphics{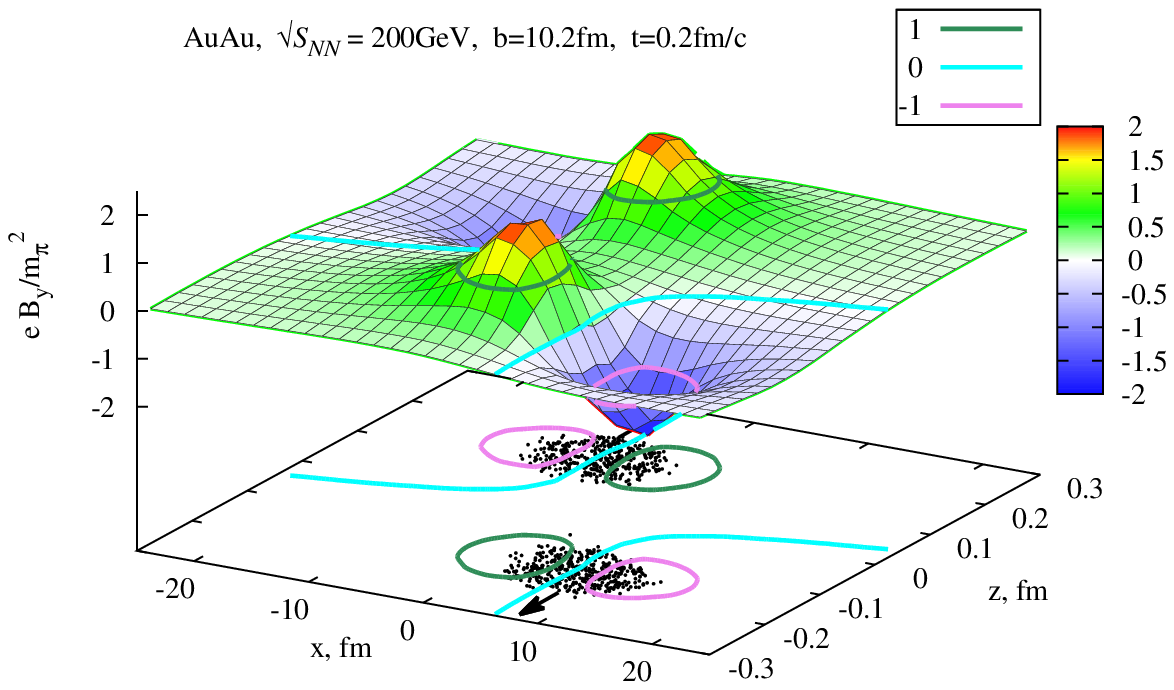} }
\resizebox{0.85\columnwidth}{!}{%
\includegraphics{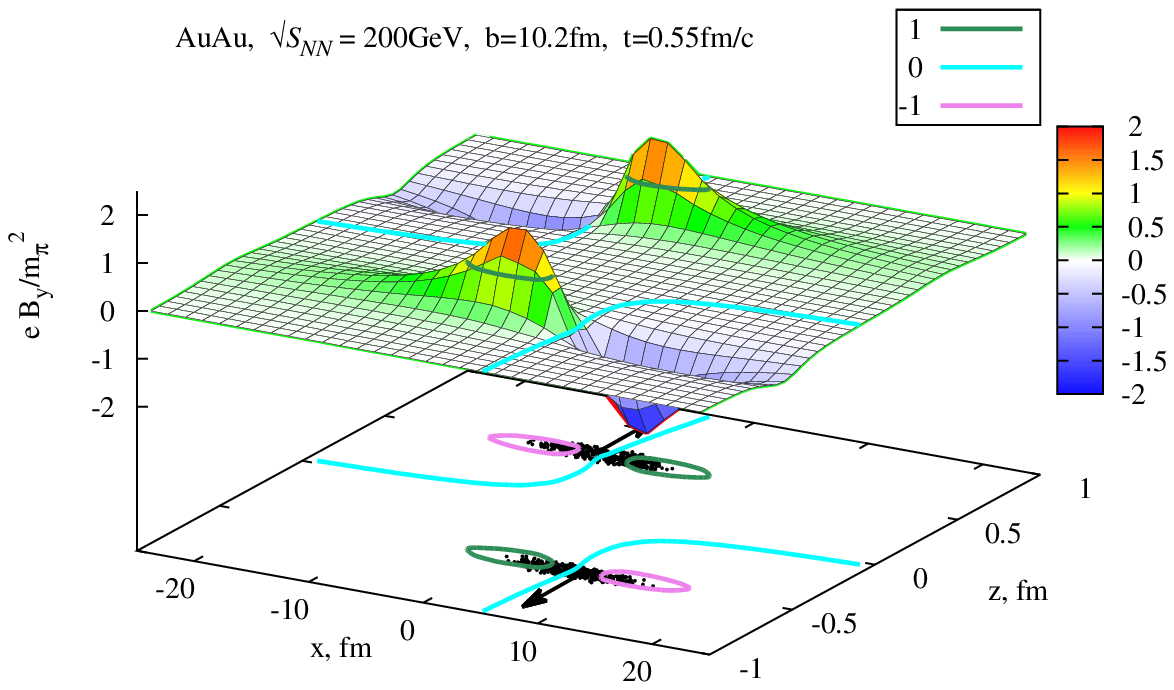} }
\caption{Distribution of the magnetic field strength $eB_y$
in the $y=0$ plane
at $t=$0.05 (upper panel), 0.2 (in the middle) and 0.55 (lower panel)
fm/c for Au+Au collisions
at $\sqrt{s_{NN}}=$200 and $b=$10.2 \ fm. The collision geometry is
projected on $x-z$ plane by points corresponding to a particular
spectator position.
Curves (and their projections) are levels of the constant $eB_y$.
  \label{By} }
\end{center}\end{figure}

One should note that the off-shell HSD transport approach 
is based not on the Boltzmann-like
transport equation (\ref{kinEq}) but rather on the off-shell Kadanoff-Baym
equations~\cite{CJ99} having similar general structure. The set of
equations was solved in a quasiparticle approximation by using the
Monte-Carlo parallel ensemble  method.
To find the magnetic field a space grid was used.
In a lattice point of this grid  the retarded vector
potential is calculated by numerical differentiation.
The field inside a cell is approximated by that at
the nearest grid point. To avoid singularities and
self-interaction effects, particles within a given
cell are excluded from  procedure
of the field calculation.

 \begin{figure}[thb]
\begin{center}
\resizebox{0.85\columnwidth}{!}{%
\includegraphics{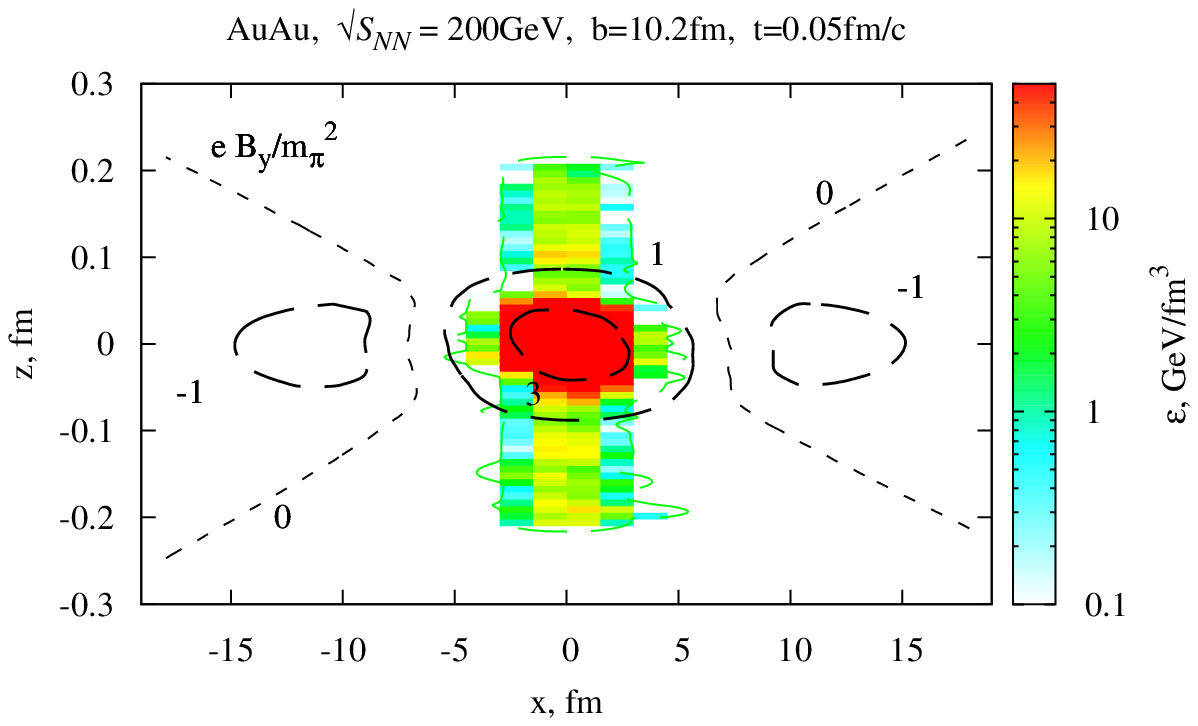} }
\resizebox{0.85\columnwidth}{!}{%
\includegraphics{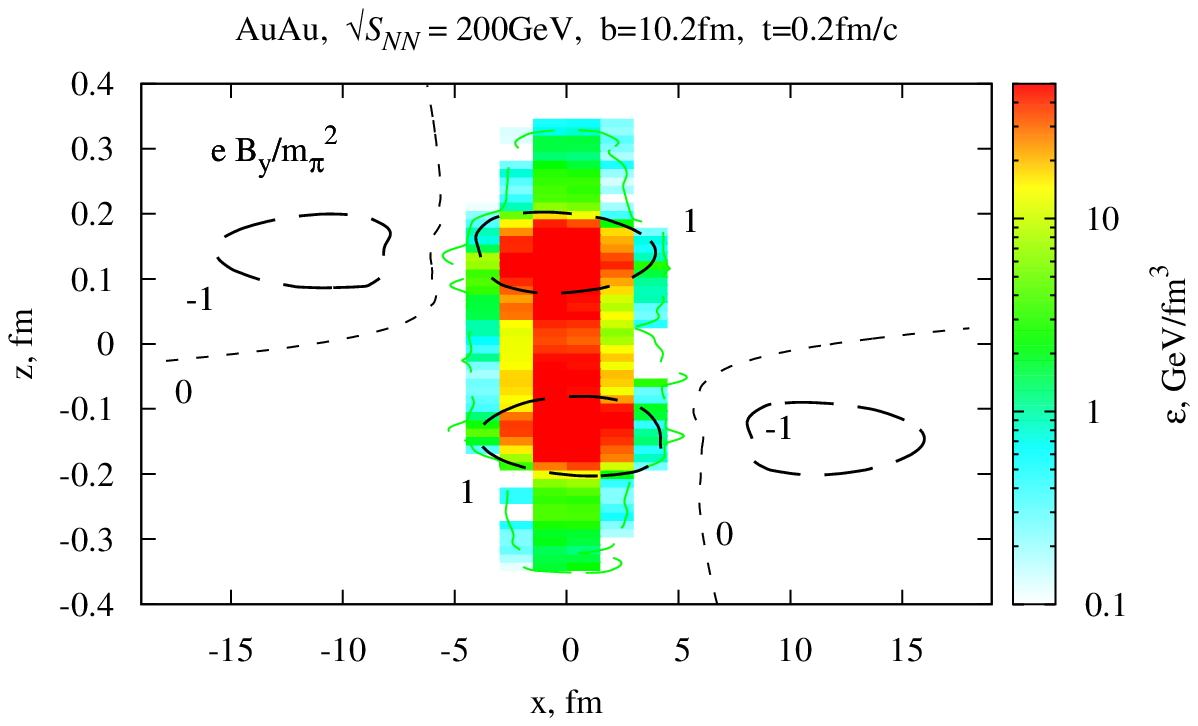} }
\resizebox{0.85\columnwidth}{!}{%
\includegraphics{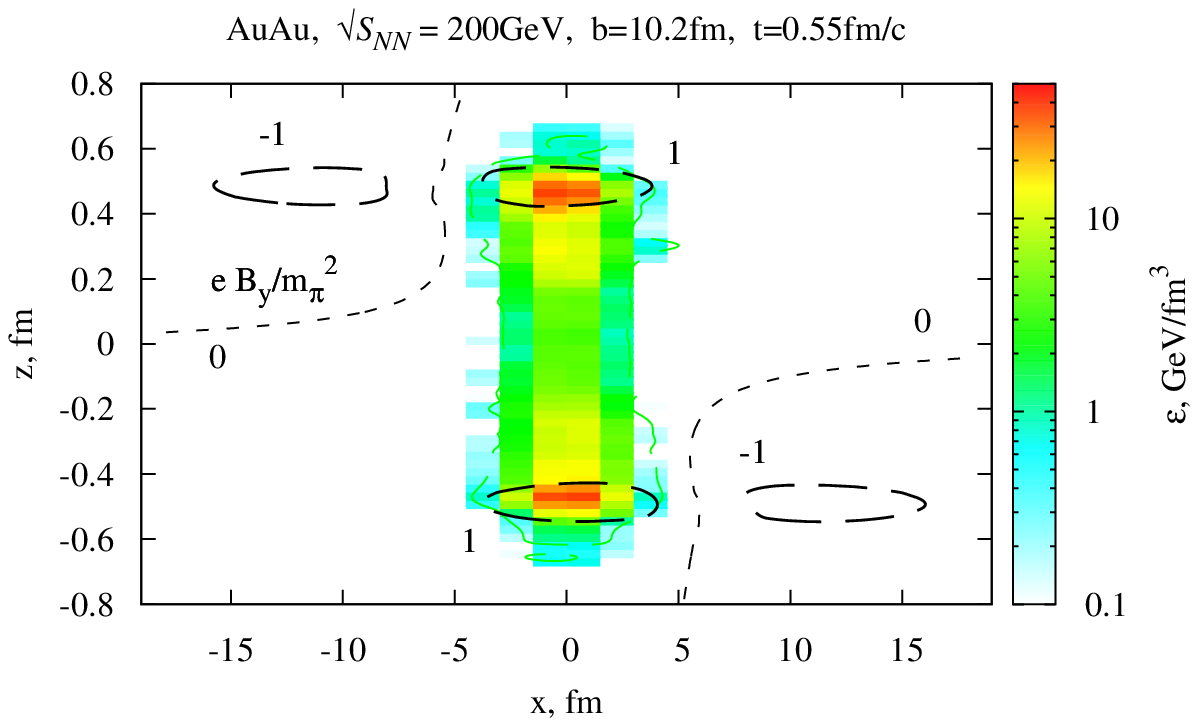} }
\caption{The $x-z$ projection of the energy density $\varepsilon$ and
magnetic field strength $eB_y$ (contour lines) in the $y=0$ plane
for three cases presented in Fig.\ref{By}.
   \label{ByEn}}
\end{center}\end{figure}

An evolution snapshot of the magnetic field $B_y(x,y=0,z,t)$ (in units
of $m_\pi^2$) formed in Au+Au (200 GeV) peripheral ($b=$10.2 fm) collisions
are given in Fig.\ref{By} for three time moments $t=$0.05, 0.20 and 0.55
fm/c. The collisional geometry is presented by a set of points every of
which corresponds to a spectator nucleons. The field is not homogeneous
exhibiting a wide maximum over the transverse size of overlapping
(participant) matter. Opposite rotation of the magnetic field along
direction of two colliding nuclei results in corresponding two minima
from outer sides of spectator matter remnants. At expansion these remnants
are moving away from each other. It is of interest that for the fixed $x$
the width of $Z$ distributions of the magnetic field does not changed much.
This type of scaling has been assumed in the phenomenological model of the
CME considered above.  The maximum in the magnetic field strongly correlates with that in the
energy density  of created particles as is clearly seen from
Fig.\ref{ByEn}. Large local values of $B_y$ and $\varepsilon$ reached
in these Au+Au collisions provide necessary conditions for  observation of
signals of a possible parity violation.

 \begin{figure}[thb]
\resizebox{0.85\columnwidth}{!}{%
\includegraphics{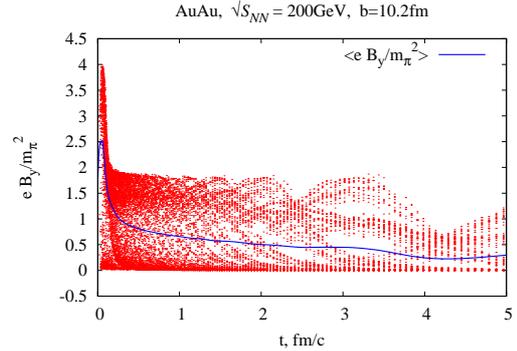} }
\caption{The time evolution of the magnetic field $eB_y$ at the
point where a pion is situated (shown by points). The curve corresponds
to the field averaged over these points at every time moment.
 \label{ByT}}
\end{figure}

A particular characteristic is shown in Fig.\ref{ByT}. The magnetic field
in the point where a real meson is situated is plotted as a function of
time for every meson. The curve corresponds to the averaged field
$<B_y/m_\pi^2>$ at the given time moment. It is seen that averaged magnetic
 field sharply goes down till $B_y<m_\pi^2$ and then falls down quite
slowly. The moment, when the expansion regime is changed, is of order of
$2R/\gamma_{cm}\sim 0.2$ fm/c.

\section{Discussion and conclusions}
Summarizing Sect.\ref{sec:2}, one should note that for
heavy-ion collisions at
$\sqrt{s_{NN}}\gsim$ 11 GeV the magnetic field and energy density
of deconfined matter reach very high values which seem to be high
enough for manifestation of the Chiral Magnetic Effect. However,
these are only necessary conditions. To estimate a possible CME
a particular model is needed.  For the average correlator  our
qualitative prediction $a^2\sim s^{-1/8}_{NN}$  has
a rather small exponent but nevertheless it is too strong to
describe observable energy behavior of
the CME. This model energy dependence can be reconcile with
experiment~\cite{Vo09} by a detailed treatment of the lifetime
taking into account both the time of being in a strong magnetic field
and time evolution of the energy density in the QGP phase. For the
chosen parameters we are able to describe data for Au+Au collisions on
charge separation at two available energies. We predict that the effect
will be much smaller at the LHC energy and will sharply disappear near
the top energy of SPS. Experiments at coming into operation Large Hadron
Collider and that planned at RHIC  by the Beam Energy Scan program
\cite{BES} are of great interest since they will allow one to test the
CME scenario and to infer the critical magnetic field
$eB_{crit}$ governing by the spontaneous local ${\cal CP}$
violation.

 All the
results presented above are obtained under assumption that
evolution of the electromagnetic field scales as that in the
central point of the overlapping region. The fall of $B_y$ in time
is caused be the fact that spectators are flying away. This
assumption is getting worse for lower SPS energies  what should be
tested in more elaborated models.

The experimentally observed CME enhancement for Cu+Cu collisions
is related with the selection of different impact parameters for
the same centrality at variation of a colliding system. However,
it is not reduced to a purely geometrical effect.

The problem of parity violation in strong interactions and the related
CME are actively debated now. In particularly, the important finding
has been reported recently~\cite{BKL09}. To
distinguish between the CME and effects due to certain two-particle
correlations it was proposed to analyze the strength of
charge separation effect and its azimuthal correlation with the
reaction plane. It turned out that the observable effect for
like-sign pions is the sum of a large negative in-plane correlator
and small positive out-of-plane one, i.e. the correlations for
same charge pairs are mainly in-plane. For unlike pions both
components are positive and approximately equal to each other. All
that is puzzling since the CME is expected only for out-of-plane
events.  A full dynamical simulation is needed for heavy-ion
collisions with inclusion of the magnetic field formation, its
evolution and its impact on particle trajectories. First step in this
direction has been made in Sec.\ref{sec:3}. Such approach allows one
to calculate the CME background. It is of great
interest that the discussed CME signal, the electric charge
asymmetry with respect to the reaction plane,  may originate not
only from the spontaneous local {\cal CP} violation but also be
simulated by other possible effects. The search for alternative
explanations  and additional manifestations of local parity
violation is
underway~\cite{BCLP09_2,Wa09,MS09,RST10,FLW10,KZO10,Ch10,Pr10,AMM10,Na10,Vol10,OS10,Zh10,MSh10,MAS10,KhS10}).
It is important that the developed kinetic approach in principle allows
one to simulate the
Chiral Magnetic effect itself. This work is in progress.

\section*{Acknowledgements}
Successful collaboration with D. Kharzeev, V. Skokov, E. Bratkovskaya,
W. Cassing, V. Konchakovski and S. Voloshin is greately acknowledged.
We are thankful to V. Koch, R. Lacey, I. Selyuzhenkov, O. Teryaev  and
J. Thomas  for comments.  V.T.  is partially
supported by the DFG grant WA 431 8-1 RUSS
and the Heisenberg-Landau grant. 
V.V. acknowledges financial support within the ``HIC for FAIR" center
of the ``LOEWE'' program.

\end{document}